# Extension of cellular automata by introducing an algorithm of recursive estimation of neighbors

Yoshihiko Kayama

BAIKA Women's University, Japan
(Tel: 81-72-643-6221, Fax: 81-72-643-8473)

kayama@baika.ac.jp

**Abstract:** This study focuses on an extended model of a standard cellular automaton (CA) that includes an extra index consisting of a radius that defines a perception area for each cell in addition to the radius defined by the CA rule. Extended standard CA rules form a sequence ordered by this index, which includes the CA rule as its first term. This extension aims at constructing a model that can be used within the CA framework to study the relationship between information processing and pattern formation in collective systems. Although the extension presented here is merely an extrapolation to a CA with a larger rule neighborhood, the extra radius can be interpreted as an individual difference of each cell, which provides a new perspective to CA. Some pattern formations in extended one-dimensional elementary CAs and two-dimensional Life-like CAs are presented. It is expected that the extended CA can be applied to various simulations of complex systems and other fields.

## 1 INTRODUCTION

Cellular automaton (CA), which consists of cells arranged in a grid, was proposed by Uram and Neuman [1] in the 1940s. In the 1970s, Conway found a two-dimensional CA, which he called "the Game of Life," that exhibited complex behaviors evoking biological activities (Gardner [2] and Berlekamp, Conway, and Guy [3]). In the 1980s, Wolfram studied one-dimensional CAs (Wolfram [4-7]). He proposed that CA could be grouped into four classes of complexity: homogeneous (class I), periodic (class II), chaotic (class III), and complex (class IV).

Under the CA framework, the neighborhood of each cell is defined by the CA rule and there is no possibility of expanding the sensory area of a cell. For example, each cell of a one-dimensional elementary CA (ECA) acquires states of the three cells within its radius-one neighborhood to determine its state in the next step. In contrast, in a flocking "boids" simulation, which is an artificial life program developed by Reynolds (Reynolds [8], and Banks, Vincent, and Anyakoha [9]), each boid obtains the motion information of other boids within its perception area and using a simple algorithm, alters its own motion according to an analysis of this information. Boids easily organize themselves into a large, orderly group and move as a single organism without a central commander, i.e., their information processing leads to a collective control of group motion. If a framework similar to that of "boids" were to be introduced into CA, we would be able to discuss the relationship between information processing and fluctuation control in CA pattern formation.

In this article, we discuss an extended model of standard CA that includes an extra index consisting of a radius that defines a perception area for each cell in addition to the radius of the rule neighborhood. Extended rules originating from a standard CA form a sequence ordered by the index, which includes the CA rule as its first term. The extended model is implemented by an algorithm named "Recursive Estimation of Neighbors" (REN), in which the original CA rule is recursively used to estimate the states of neighboring cells. Typical examples of extended CA originating from one- and two-dimensional binary CAs are presented.

The next section shows how the radius of perception area of a cell can be introduced in addition to the radius of the CA neighborhood by using the REN algorithm. In Section 3, the definition of the extended CA is applied to ECA and to two-dimensional eight-neighbor outer-totalistic CA including Conway's Game of Life. With the additional radius, some sequences of the extended ECA change complexity between periodic and chaotic patterns. In the sequence of the extended Game of Life, there is a positive correlation between the presence of the additional radius and the average convergence speed to a rest state; for another sample, there is a negative correlation. Some interesting pattern formations are also found in heterogeneous models; that is, their grids are composed of cells that follow different extended rules over a given sequence.

## 2 EXTENSION OF CA

In order to expand the perception area of a cell, it should be separated from the neighborhood determined by the CA rule. In case of boids, each boid acquires information regarding the positions and velocities of boids within its perception area and determines its own movement so as to follow the representative values of the neighbors. A similar scenario can be written in CA: in applying the CA rule, each cell acquires the states of cells within its perception area and considers virtual cells with representative values, e.g., average values of this information, to be its neighbors. In the boids case, the approach of applying the update process to the representative values controls fluctuation and leads to a group formation; in extended CA, by contrast, the result of this scenario is expected to simply resemble a coarse graining of the grid. Because no notable results have been obtained so far using this approach, further arguments will not be provided for this scenario.

We then try to explore a model with an expanded perception area that includes the rule neighborhood. The original CA rule is referred to as the "base rule." The target model must function without introducing any other rules and must be able to treat the size of the perception area as a parameter expressing differences between individual cells. In other words, we take the standpoints that pattern formations can be discussed on the basis of a possible single architecture and that some complexity of pattern formation can emerge from differences between individual elements.

To illustrate the basic concept of separation of the perception area from the rule neighborhood, we consider our decision-making as to what each cell is to do in the next step. This action will be accompanied by a reasoning process, i.e., the action will be decided through prediction of neighbors' actions from current information. One of the rules of boids— "Alignment," which dictates that each boid follows the average speed of the boids within its perception area— contains a reasoning process. Here, we propose an algorithm of REN, in which the base CA rule is used recursively to estimate the neighbors' states in the next step until the state of the target cell is subsequently determined.

The following assumptions will make the definition of the extended CA clear:

(i) Each cell has a perception area in addition to the neighborhood defined by the base CA rule. The states of all cells within the area are perceived in each step.
(ii) In each step, a cell determines its own state in the next step by estimating the states of the cells within its perception area in the next step, using the base rule recursively.
(iii) In the process of estimation, the same algorithm of recursive estimation is assumed to be used by all cells within the perception area. Within the perception area, each cell is assumed to have a perception area that is as large as possible.
(iv) In cases where the assumed perception area of a cell is smaller than the neighborhood of the base rule, the base rule is not applicable and the state of the cell in the next step is assumed to remain equal to the current state.

In the following, we assume that the rule neighborhood and perception area are both isotropic and can be parametrized by their respective radii $r$ and $R$. Their containment relationship is expressed by $R \geq r$. Here, we restrict our discussion to the simplest one-dimensional binary CA with $r = 1$, called ECA. The state of the $i$-th cell at time $t$ and an ECA rule function are denoted by $x_i^{(t)}$ and $f$, respectively. The time evolution of the state is expressed by

$$x_i^{(t+1)} = f(x_{i-1}^{(t)}, x_i^{(t)}, x_{i+1}^{(t)}). \quad (1)$$

In the REN framework, the above equation changes to

$$\varphi_{R,i}^{(t+1)} = f(\varphi_{R-1,i-1}^{(t+1)}, x_i^{(t)}, \varphi_{R-1,i+1}^{(t+1)}), \quad (2)$$

where $\varphi_{R,i}^{(t+1)}$ is an estimated state of the $i$-th cell with radius $R$ at $t+1$ and $\varphi_{R-1,i\pm1}^{(t+1)}$ indicate the estimated states of the neighbors at $t+1$ with an estimated radius $R$-1; the value of the neighbors' radius stems from assumption (iii) because $R$-1 is the maximum value of the perception area for neighbors within the perception area of the $i$-th cell. Note that $\varphi_{R,i}^{(t+1)}$ is equal to the actual state $x_i^{(t+1)}$, but $\varphi_{R-1,i\pm1}^{(t+1)}$ are not necessarily equal to their respective actual states $x_{i\pm1}^{(t+1)}$ because the neighbors' true radii are not $R-1$ but $R$. Subsequently, assumption (iii) leads to the following recursive expressions of estimated states of the neighbors:

$$\begin{aligned}\varphi_{R-j,i-j}^{(t+1)} &= f(\varphi_{R-j-1,i-j-1}^{(t+1)}, x_{i-j}^{(t)}, \varphi_{R-j-1,i-j+1}^{(t+1)}), \\ \varphi_{R-j,i+j}^{(t+1)} &= f(\varphi_{R-j-1,i+j-1}^{(t+1)}, x_{i+j}^{(t)}, \varphi_{R-j-1,i+j+1}^{(t+1)}),\end{aligned} \quad (3)$$

where $j = 1, 2, \ldots, R-1$. Because $j = R$ implies that none of the neighbors' perception areas are estimated, assumption (iv) gives the following conditions:

$$\varphi_{0,i\pm R}^{(t+1)} = x_{i\pm R}^{(t)}, \quad \varphi_{0,i\pm R\mp2}^{(t+1)} = x_{i\pm R\mp2}^{(t)}. \quad (4)$$

As the first step of a concrete demonstration, let us consider the case $R = r = 1$. Eqs. (2) and (4) give

$$x_i^{(t+1)} = f(\varphi_{0,i-1}^{(t+1)}, x_i^{(t)}, \varphi_{0,i+1}^{(t+1)})$$

and $\varphi_{0,i\pm1}^{(t+1)} = x_{i\pm1}^{(t)}$. Accordingly, the extended ECA with $R = 1$ is identical to the base ECA (eq. (1)). This discussion is not restricted to ECA; all extended CAs with $R = r$ are identical to their base CA.

We next discuss the case $R = 2$, which corresponds to a 5-neighbor rule. Eq. (2) gives
$$x_i^{(t+1)} = f(\varphi_{1,i-1}^{(t+1)}, x_i^{(t)}, \varphi_{1,i+1}^{(t+1)}),$$
and the recursive expressions (3) give
$$\begin{aligned}\varphi_{1,i-1}^{(t+1)} &= f(\varphi_{0,i-2}^{(t+1)}, x_{i-1}^{(t)}, \varphi_{0,i}^{(t+1)}),\\ \varphi_{1,i+1}^{(t+1)} &= f(\varphi_{0,i}^{(t+1)}, x_{i+1}^{(t)}, \varphi_{0,i+2}^{(t+1)}),\end{aligned}$$
Because from eqs. (4), $\varphi_{0,i}^{(t+1)} = x_i^{(t)}$ and $\varphi_{0,i\pm2}^{(t+1)} = x_{i\pm2}^{(t)}$, the extended $R = 2$ CA is expressed by
$$x_i^{(t+1)} = f\left(f(x_{i-2}^{(t)}, x_{i-1}^{(t)}, x_i^{(t)}), x_i^{(t)}, f(x_i^{(t)}, x_{i+1}^{(t)}, x_{i+2}^{(t)})\right). \quad (5)$$
The cases for larger values of $R$ can be derived similarly.

The above discussion indicates that the extended rules with increasing values of $R$ form a sequence parametrized by $R$ and originating from the base CA rule. As each extended rule for $R$ is a $(2 \times R + 1)$-neighbor CA rule, this extension represents an extrapolation via the REM algorithm to CA with increasing neighborhood sizes.

## 3 PATTERN FORMATIONS

A treatment similar to the extension discussed in the previous chapter can be applied to the two-dimensional outer-totalistic CA, including Conway's Game of Life (called Life-like CA). Here, some examples of pattern formations in extended ECA and Life-like CA are presented.

### 3.1 Extended ECA

ECA is the simplest nontrivial CA with $r = 1$; its $2^3 = 8$ different neighborhood configurations result in $2^8 = 256$ possible rules. We follow a standard naming convention invented by Wolfram [4, 7], which gives each rule in ECA a number from #0 to #255. The equivalency of the CA rules under mirror and complement transformations reduces the number of independent rules to 88 (Li and Packard [10], and Kayama [11]).

A sequence of extended rules with index $R$ is represented by a code in which the base rule is enclosed by square brackets, e.g., [#110]. When each rule included in the sequence is identified, the rule code is shown as the base rule code followed by a letter "$R$" and its value. The sequence is represented by
$$[\#110] = \{\#110R1, \#110R2, \#110R3, \dots\},$$
where #110R1 is identical with the base rule #110. In the simulations used in this subsection, we set the maximum value of $R$ to 20.

Among the sequences generated from independent ECA rules, eight are based on class I rules and all of the rules contained in these sequences also belong to class I. In contrast, various pattern formations can be found in sequences based on class II rules. The sequence [#134] shows changes between periodic and chaotic patterns depending on the index $R$ (Fig. 1), where the green dots are live cells and black ones are dead. The patterns originate from initial configurations with one live cell. Some sequences other than [#134] show similar changes.

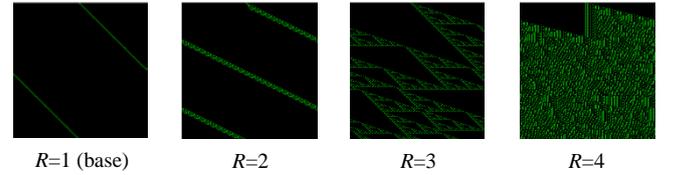

$R=1$ (base)    $R=2$    $R=3$    $R=4$
**Figure 1** Pattern formations in [#134]

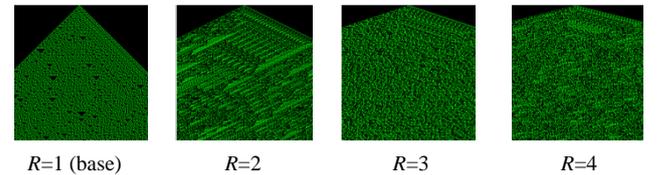

$R=1$ (base)    $R=2$    $R=3$    $R=4$
**Figure 2** Pattern formations in [#30]

Pattern formations in sequences based on class III and IV rules are also attractive. Class III rules are sometimes exemplified by rule #30, and its sequence [#30] has typical chaotic patterns (Fig. 2). Some sequences show periodic changes between periodic and chaotic patterns depending on the odd-even parity of $R$, e.g., [#22] (Fig. 3) and [#110] (Fig. 4). In these cases, no simple correlations have been found between fluctuation control and the radius $R$, even when the amount of information each cell acquires increases monotonically with $R$.

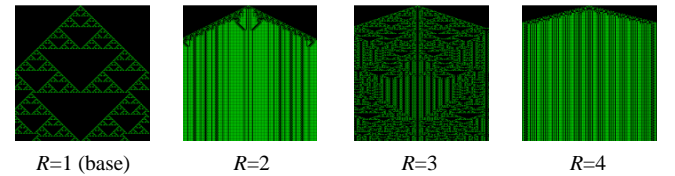

$R=1$ (base)    $R=2$    $R=3$    $R=4$
**Figure 3** Pattern formations in [#22]

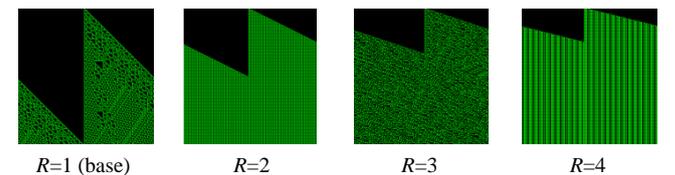

$R=1$ (base)    $R=2$    $R=3$    $R=4$
**Figure 4** Pattern formations in [#110]

### 3.2 Extended Life-like CA

In the descriptions below, all two-dimensional Life-like CA rules are specified in the Golly/RLE format (Adamatzky (Ed.) [12], and Eppstein [13]). Conway's

Game of Life is denoted by B3S23 in this notation, where "B" stands for "birth" and "S" stands for "survival." Following the previous subsection, a sequence of extended rules is denoted by the base rule code in square brackets, e.g., [B3S23].

In Conway's Game of Life, one of the most famous CA, many complex patterns and activities can emerge (Callahan [14], Flammenkamp [15]). After a long transient process, a randomly generated initial configuration transfers to a rest state that can include a variety of patterns: still lifes (e.g., blocks, beehives, or ships) and oscillators (e.g., blinkers, toads, or beacons). Isolated still lifes are also still lifes in any extended rule in the sequence [B3S23] based on the Game of Life because the estimated states $\varphi_{R,i}^{(t+1)}$ defined in Chapter 2 are all equal to the actual states $x_i^{(t+1)}$. Figure 5 shows rest states of [B3S23], where colored dots represent live cells. The average transient time from pseudo-randomly generated initial configurations to rest states decreases (Fig. 6), which means that there is a positive correlation between the radius $R$ and the average convergence speed. The long transient time at $R = 6$ is the only exception owing to a glider (Fig. 7-b travels diagonally). Additionally, an amusing glider can be found in B3S23R2 (Fig. 7-a travels vertically).

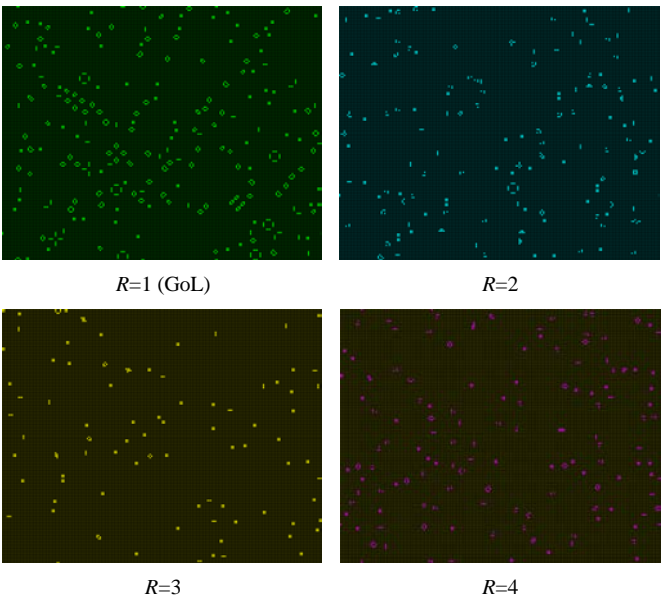

**Figure 5** Rest states of [B3S23]

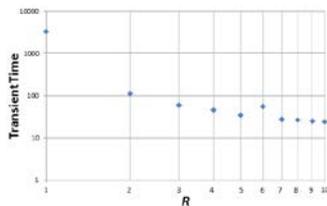

**Figure 6** Log-log plots of transient times in [B3S23]

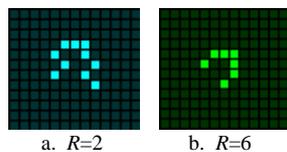

a. $R=2$    b. $R=6$
**Figure 7** Gliders in [B3S23]

In contrast, [B23S234] has a rest state only in B23S234R1 and random states in all others (Fig. 8). Randomness gradually increases with $R$.

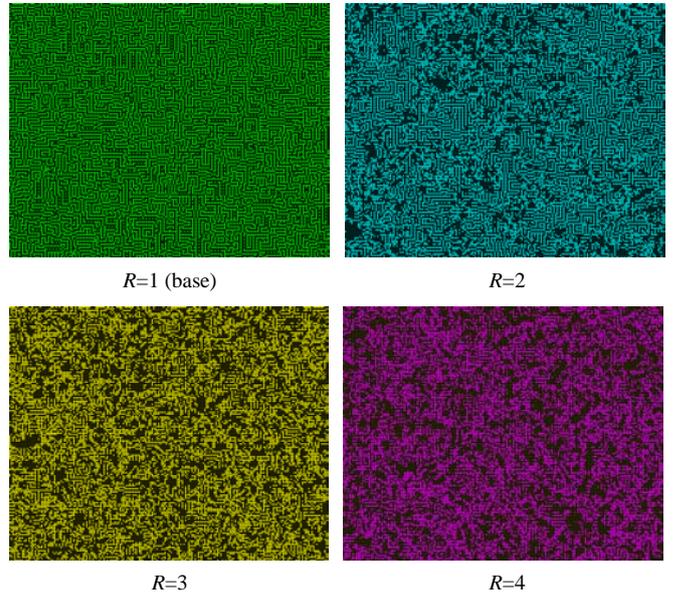

**Figure 8** Pattern formations in [B23S234]

Thus far, we have discussed homogeneous CA, that is, all cells following the same CA rule. If we recognize the extra radius $R$ as an index of the amount of information each cell can acquire, heterogeneous CA composed of cells having different values of $R$ becomes meaningful. Figs. 9-a and -b represent mixing between B3S23R1 and B3S23R2 cells and between B23S234R3 and B23S234R4 cells, respectively. The mixing ratio changes linearly from 1:0 (left-hand side) to 0:1 (right-hand side). Although B3S23R1 and B3S23R2 have rest states as shown in Fig. 5, Fig. 9-a shows that their intermediate mixing area is unstable. In contrast, Fig. 9-b shows the emergence of a rest state in the intermediate area. Nevertheless, B23S234R3 and B23S234R4 have no rest states (Fig. 8).

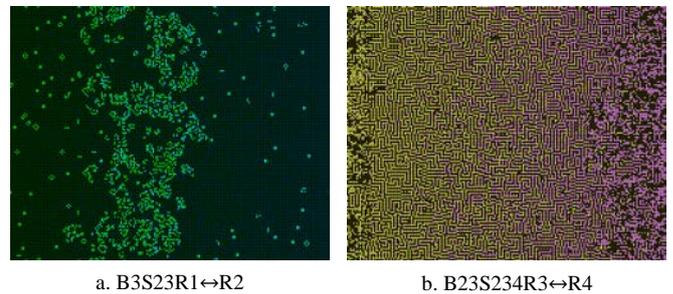

a. B3S23R1↔R2    b. B23S234R3↔R4
**Figure 9** Mixings of two types of cells

A mixing of B4S1234R1 and B4S1234R2 cells is also interesting. Fig. 10 shows a complex pattern change; there is an area of islands between two walls and a random state at the right-hand edge. The left-hand edge wall is a rest state of B4S1234R1 and the right-hand edge is a random state of B4S1234R2. Their mixing ratio is the same as in the case of Fig. 9, but the initial probability of live cells has been set to 0.2. Some analytical discussion to better understand the above results should be presented in future work.

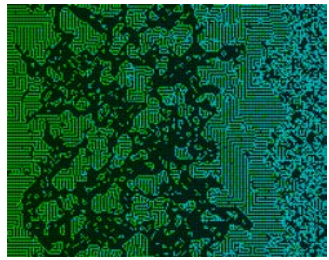

**Figure 10** B4S1234R1↔R2

## 4 CONCLUSIONS

In this article, we tried to extend the standard CA to a model that can be applied to studying the relationship between information processing and pattern formation in collective systems with a minimal introduction of extra rules or technical contrivances. The REN algorithm, which arose from the idea of self-similarity, made it possible to separate the perception area from the rule neighborhood and to formalize an extension of the rule. There may be other algorithms besides REN to estimate the states of neighbors, and the assessment of possible candidates is a future subject of investigation. In the applications discussed in this paper, the extended ECA sequences showed a variety of patterns that changed with the size of the perception area of each element. The sequences of extended Life-like CA, [B3S23] and [B23S234], represent contrasting dependences on the size of the perception area. It will be crucial to understand these differing correlations in order to construct an applicable model for various simulations of complex systems and other fields.

Heterogeneous models of the extended CA show further possibility. Even if the homogeneous models have rest states, a heterogeneous model of cells belonging to the various homogeneous models within a given extended sequence may have unstable patterns, and vice versa. Such phenomena appear to act like a mixture of two different materials changing its state through a chemical reaction. Just as the boids theory has been adapted to "Swarm Chemistry" by introducing interactions between its elements and an evolutionary algorithm (Sayama [16-18]), the extended CA would be expected to evolve into "CA Chemistry."

## REFERENCES

[1] Neumann J. von (1966), The theory of self-reproducing automata. In: Burks A. W. (Ed.), Essays on Cellular Automata, University of Illinois Press

[2] Gardner M. (1970), Mathematical games. Scientific American, vol. 223, pp. 102-123

[3] Berlekamp E. R., Conway J. H., and Guy R. K. (1982), Winning Ways for Your Mathematical Plays. Academic, New York

[4] Wolfram S. (1983), Statistical mechanics of cellular automata. Reviews of Modern Physics, vol. 55, pp. 601-644

[5] Wolfram S. (1984), Universality and complexity in cellular automata. Physica D, vol. 10, pp. 1-35

[6] Wolfram S. (1986), Theory and Applications of Cellular Automata. World Scientific, Singapore

[7] Wolfram S. (2002), A New Kind of Science. Wolfram Media, Inc.

[8] Reynolds C. W. (1987), Flocks, herds and schools: A distributed behavioral model. ACM Siggraph Computer Graphics, vol. 21 (4), pp. 25-34

[9] Banks A., Vincent J., and Anyakoha C. (2007), A review of particle swarm optimization. Part i: background and development. Natural Computing, vol. 6 (4), pp. 467-484

[10] Li W., Packard N. (1990), The Structure of the Elementary Cellular Automata Rule Space, Complex Systems, vol. 4, pp. 281–297

[11] Kayama Y. (2011), Network Representation of Cellular Automata. In 2011 IEEE Symposium on Artificial Life (IEEE ALIFE 2011) at SSCI 2011, pp. 194-202

[12] Adamatzky A. (Ed.) (2010), Game of life cellular automata. London, Springer

[13] Eppstein D. (2010), Growth and decay in Life-like cellular automata. In Adamatzky A. (Ed.), Game of Life Cellular Automata, Springer, pp. 71–98

[14] Callahan P. (1995), Patterns, Programs, and Links for Conway's Game of Life. http://www.radicaleye.com/lifepage/. Retrieved at February 1, 2011

[15] Flammenkamp A. (1998), Achim's Game of Life. http://wwwhomes.uni-bielefeld.de/achim/gol.html Retrieved at December 12, 2011

[16] Sayama H. (2007), Decentralized control and interactive design methods for large-scale heterogeneous self-organizing swarms. Advances in Artificial Life, vol. 15(1), pp. 105-114

[17] Sayama H. (2009), Swarm chemistry. Artificial Life, vol. 15(1), pp. 105-114

[18] Sayama H. (2010), Robust morphogenesis of robotic swarms. Computational Intelligence Magazine, IEEE, vol. 5(3), pp. 43-49